\documentclass[fleqn,10pt]{wlscirep}

\usepackage[utf8]{inputenc}
\usepackage{graphicx,epsfig}
\usepackage{times}
\usepackage{mathtools,amssymb,amsmath,amsfonts,multirow,rotate,color}
\usepackage[caption=false]{subfig}
\usepackage{color}
\usepackage{soul}
\usepackage{tikz}
\usepackage{verbatim}
\usepackage{pgfplots}
\usepackage{appendix}
\usepackage{mathrsfs}
\usepackage{wasysym}
\usepackage{stmaryrd}
\usepackage{chngcntr}
\usepackage{algpseudocode,algorithm}

\usetikzlibrary{patterns}
\usetikzlibrary{positioning}

\pgfplotsset{compat=1.15}

\title{Rank the spreading influence of nodes using dynamic Markov process}

\author[1,2,3]{Jian-Hong Lin}
\author[4]{Zhao Yang}
\author[5]{Jian-Guo Liu\thanks{Corresponding author: liujg004@ustc.edu.cn}}
\author[6,7]{Bo-Lun Chen\thanks{Corresponding author: chenbolun@hyit.edu.cn}}
\author[1,2]{Claudio J. Tessone\thanks{Corresponding author: tessone@ifi.uzh.ch}}

\affil[1]{Blockchain and Distributed Ledger Technologies, Institute of Informatics, University of Z\"urich, Andreasstrasse 15, CH-8050 Z\"urich (Switzerland)}
\affil[2]{UZH Blockchain Center, University of Z\"urich, Andreasstrasse 15, CH-8050 Z\"urich (Switzerland)}
\affil[3]{ETH Z\"urich, Department of Management, Technology and Economics, Scheuchzerstrasse 7, CH-8092 Z\"urich (Switzerland)}
\affil[4]{Global People Analytics, Talent Rewards $\&$ Insights, Roche, CH-4070 Basel (Switzerland)} 
\affil[5]{Institute of Accounting and Finance, Shanghai University of Finance and Economics, Shanghai 200433, PR China}
\affil[6]{Faculty of Computer and Software Engineering, Huaiyin Institute of Technology, Huaian 233003, PR China}
\affil[7]{Swiss Center for Data and Network Sciences, University of Zurich, CH-8050 Z\"urich, (Switzerland)}

\begin{abstract}
Ranking the spreading influence of nodes is of great importance in practice and research. The key to ranking a node’s spreading ability is to evaluate the fraction of susceptible nodes been infected by the target node during the outbreak, i.e., the outbreak size. In this paper, we present a dynamic Markov process (DMP) method by integrating the Markov chain and the spreading process to evaluate the outbreak size of the initial spreader. Following the idea of the Markov process, this method solves the problem of nonlinear coupling by adjusting the state transition matrix and evaluating the probability of the susceptible node being infected by its infected neighbours. We have employed the susceptible-infected-recovered (SIR) and susceptible-infected-susceptible (SIS) models to test this method on real-world static and temporal networks. Our results indicate that the DMP method could evaluate the nodes’ outbreak sizes more accurately than previous methods for both single and multi-spreaders. Besides, it can also be employed to rank the influence of nodes accurately during the spreading process.

\end{abstract}

\begin{document}

\maketitle

\thispagestyle{empty}

\section{Introduction}
Complex networks are widely used to represent interactions between people, technology, and various entities. Among all the studies within the area of network theory, understanding the dynamics of spreading processes is of particular interest. Although the spreading dynamics on networks are not a new phenomenon, studies in this field lead to better understandings of many important social and natural processes~\cite{tao2006epidemic}, such as the spreading of infectious diseases~\cite{keeling2008modeling}, the propagation of computer virus~\cite{kephart1997fighting},  the cascading process~\cite{motter2004cascade}, traffic congestion~\cite{li2015percolation},
the centralization in Bitcoin system\cite{lin2020lightning,lin2021weighted}, and so on. One important approach in studying the spreading dynamics is to estimate and rank nodes' spreading abilities. Through this approach, one might first locate influential nodes of complex networks and later on control the outbreak of epidemics ~\cite{pastor2002immunization,cohen2003efficient,chen2022graph}, target the opinion leaders in social networks ~\cite{bodendorf2009detecting,uehara2019analysis}, quantify the scientific impact~\cite{sinatra2016quantifying,wang2013quantifying}, and accelerate the adoption of innovation~\cite{pei2013spreading}, etc.

Classical centrality measures have been developed to identify the nodes' spreading influence. The degree centrality~\cite{freeman1978centrality} is probably the most straightforward one. Nodes with larger degree centrality are considered to have better spreading abilities than the other nodes within a graph. The betweenness centrality~\cite{freeman1977set}, which calculates the number of shortest paths cross through a certain node, represents the controllability of information flow over the networks. The closeness centrality~\cite{sabidussi1966centrality} measures the inverse of the mean geodesic distance from a certain node to all other nodes. The more central a node is, the closer it is to all the other nodes.  The eigenvector centrality~\cite{bonacich2001eigenvector} assigns relative scores to all nodes in the network based on the concepts that connections to influential nodes, i.e., high-scoring nodes, would be more important than that to low-scoring nodes. The $k$-shell decomposition method~\cite{kitsak2010identification} assigns nodes to different shells and considers those located within the core of the network are the most efficient spreaders. Furthermore, a lot of methods for identifying the node spreading influence have been developed from different perspectives ~\cite{liu2013ranking,ren2014iterative,lin2014identifying,lu2016h,alvarez2015eigencentrality,lawyer2015understanding}

The classical centrality measures are based on the network topological structure solely. However, recent studies have shown that the nodes' spreading influence is determined not only by the network structure but also the parameters of the dynamical processes. Therefore, various structural based centralities cannot properly identify nodes' influences since the rankings remain the same under different dynamical parameters. \u{S}iki\'{c} \emph{et al.}~\cite{vsikic2013epidemic} argued that for a given susceptible-infected-recovered (SIR) model~\cite{hethcote2000mathematics}, the rank of nodes' influence largely depends on the spreading rate and recovering rate. Klemn \emph{et al.}~\cite{klemm2012measure} suggested that the eigenvector centrality could only identify the nodes' spreading influence accurately when the spreading rate is close to the inverse of the largest eigenvalue of the network~\cite{hethcote2000mathematics}. Considering the susceptible-infected-susceptible (SIS) model~\cite{pastor2001epidemic}, Ide \emph{et al.}~\cite{ide2013diffusion} have proposed a numerical framework that uses the importance of the centrality type to determine how the vulnerable nodes change along the diffusion phases. Besides, Liu \emph{et al.}~\cite{liu2016locating} have described the infectious probabilities of nodes by a matrix differential function and have developed the dynamics-sensitive (DS) centrality to predict the outbreak size for ranking nodes' spreading influence.

The centrality measures proposed in~\cite{vsikic2013epidemic,klemm2012measure,liu2016locating} are all linear methods based on discrete Markov process. However, it is important to note that the spreading process in SIR and SIS model is usually non-linear couple process. Therefore, without taking the non-linear couple process into consideration, all the nodes' influence would be over estimated. For instance, if a susceptible node has $n^{'}$ infected nodes, the probability of this node to be infected is $1-(1-\beta)^{n^{'}}$ instead of $n^{'}\beta$ approximated by the linear methods, where $\beta$ is the spreading rate in the SIR or SIS model. 
In this paper, we present a dynamic Markov process (DMP) to evaluate the outbreak size of the nodes at given time steps. This method can be directly applied in ranking nodes' spreading influence. It overcomes the problem of nonlinear coupling by calculating the susceptible node to be infected by its neighbours sequentially and adjusting the state transition matrix during the spreading process. Our simulation results on susceptible-infected-recovered (SIR) model show that the DMP method has comparable accuracy to the linear methods~\cite{vsikic2013epidemic,klemm2012measure,liu2016locating} for both single spreader and multi-spreaders~\cite{chen2022influence}. Furthermore, we have employed the susceptible-infected-recovered (SIR) and susceptible-infected-susceptible (SIS) models to test the DMP method on real static and dynamic networks~\cite{hethcote2000mathematics, pastor2001epidemic,Scholtes2014C,Pfitzner2013B}. The simulation results show that the DMP method can rank nodes' spreading influence accurately.

\section{Methods}
\subsection{Centrality Measures.}

A network $G=(V,E)$ with $n=|V|$ nodes and $e=|E|$ links could be described by an adjacency matrix $\mathbf{A}=\left\{a_{ij}\right\}$ where $a_{ij}=1$ if node $i$ is connected to node $j$, and $a_{ij}=0$ otherwise. For directed network, if only node $i$ is pointing to node $j$, then $a_{ij}=1$ and $a_{ji}=0$.

The degree of node $i$ is defined as the number of its neighbors, namely
\begin{equation}
k_{i}=\sum_{j=1}^{n}{a_{ij}},
\label{equation2}
\end{equation}
where $a_{ij}$ is the element of matrix $\mathbf{A}$.

The main idea of eigenvector centrality is that a node's importance is not only determined by itself, but also by its neighbours' importance ~\cite{bonacich2001eigenvector}. Accordingly, eigenvector centrality of node $i$, i.e. $v_{i}$, is defined as
\begin{equation}
v_{i}=\frac{1}{\lambda}\sum_{j=1}^{n}{a_{ij}v_{j}},
\label{equation3}
\end{equation}
where $\lambda$ is a constant. Obviously, Eq.~\ref{equation3} can be written in a compact form as
\begin{equation}
\mathbf{A}\textbf{v}=\lambda\textbf{v},
\label{equation4}
\end{equation}
where $\textbf{v}=(v_1,v_2,\cdots,v_n)^T$. That is to say, $\textbf{v}$ is the eigenvector of the adjacency matrix $\mathbf{A}$ and $\lambda$ is the corresponding eigenvalue. According to Perron-Frobenius Theorem \cite{hom1985matrix}, the elements of the leading eigenvector are positive. Since the influences of nodes should be positive, $\textbf{v}$ must be the leading eigenvector corresponding to the largest eigenvalue of $\textbf{A}$, therefore we have $\textbf{v}=\textbf{q}_{1}$.

\subsection{Dynamic Markov Process Method}
In order to calculate the probabilities of nodes to be infected, one needs to solve the problem of nonlinear couple during the spreading process. Consider the union of a finite number of event $A^{'}_{1},\dots,A'_{n'}$, the probability of the event $\cup{_{i=1}^{n'}}A_{i}'$ could be written as~\cite{DeGroot2012}: 
\begin{equation}
Pr\cup{_{i=1}^{n'}}A_{i}'=1-\prod_{i=1}^{n^{'}}(1-Pr(A^{'}_{i}))=Pr(A^{'}_{1})+(1-Pr(A^{'}_{1}))Pr(A^{'}_{2})+\dots+\prod_{i=1}^{n^{'}-1}(1-Pr(A^{'}_{i}))Pr(A^{'}_{n^{'}}).
\label{equation5}
\end{equation}


The Eq.~\ref{equation5} means that the probability of the event $\cup{_{i=1}^{n'}}A_{i}'$ equals to the probability of the event $A_1^{'}$ plus the probability of the event $A_2^{'}$ while event $A_1^{'}$ doesn't happen plus the probability of the event $A_3^{'}$ while both event $A_1^{'}$ and $A_2^{'}$ do not happen, so on and so forth. Based on the above-described probability theorem, for a susceptible node with $n^{'}$ infected neighbors, the probability of being infected during the spreading process can be represented in the non-linear format as $1-(1-\beta)^{n^{'}}$, or as $\beta+(1-\beta)\beta+(1-\beta)^{2}\beta+\dots+(1-\beta)^{n^{'}-1}\beta$, where the first term $\beta$ represents the probability that this node has been infected by its first infected neighbor, the second term $(1-\beta)\beta$ represents the probability that it has not been infected by its first infected neighbor but has been infected by its second infected neighbor, and the third one $(1-\beta)^2\beta$ represents the probability that the node has not been infected by its first infected neighbor nor its second infected neighbor but has been infected by its third infected neighbor, etc.

By combining the above-described process with the standard SIR model where an infected node would infect its susceptible neighbors with a spreading rate $\beta$ and recover immediately, we propose a dynamics Markov process method as follows: Define $\textbf{x}(t)$ ($t\geq 0$) as an $n\times1$ vector whose components are approximated as the probabilities of nodes to be infected at time step $t$. Especially, if node $i$ is the initially infected node, then $x_{i}(0)=1$ and $x_{j\neq{i}}(0)=0$. In the dynamics Markov process, the initial Markov state transition matrix $\mathbf{M}=\mathbf{A}^{T}$, where $\mathbf{A}^{T}$ is the transpose of $\mathbf{A}$. If $m_{ij}=0$, node $j$ could not be infected by node $i$ anymore. Otherwise $m_{ij}$ is the probability of node $i$ to be susceptible. When $t=0$, if node $i$ is the initial infected node, it could not be susceptible anymore. Therefore we have $m_{ij}=0$, where $j=1,2,\dots n$. When $t\geq1$, we denote $\mathbf{C}(t)$ as an $n\times{n}$ matrix, where $c_{ji}(t)$ is the probability of node $j$ to be infected by node $i$ at time step $t$.

The updating rules are described as below. We first calculate the influence of node $1$ to all its susceptible neighbors. If $x_{1}(t-1)>0$, node $1$ would infect its susceptible neighbors with a probability $\beta$ at time step $t$. The probability of node $j$ to be infected by node $1$ is then
\begin{equation}
c_{j1}(t)=\beta m_{j1}x_1(t-1),
\label{equation6}
\end{equation}
where $j=1,2,\dots n$. After that, for all the nodes $j$, if $m_{jl}>0$, where $l=1,2,\dots n$, we calculate the probability of node $j$ to be susceptible, i.e. the probability of node $j$ that has not been infected by node $1$. It equals to $m_{jl}-c_{j1}(t)$. After that, for all the nodes $j$, we update the element of the state transition matrix as follows:

\begin{equation}
m_{jl}:=m_{jl}-c_{j1}(t),
\label{equation7}
\end{equation}
where $l=1,2,\dots n$. Once the state transition matrix has been updated, we continue to calculate the impact of node $2, 3, 4, ... n$ sequentially in the same way. In the end, the probabilities of node $i$ been infected at time step $t$ is
\begin{equation}
x_{i}(t)=\sum_{j=1}^{n}c_{ij},
\label{equation8}
\end{equation}

\begin{figure}[t!]
\begin{center}
\scalebox{0.72}[0.72]{\includegraphics{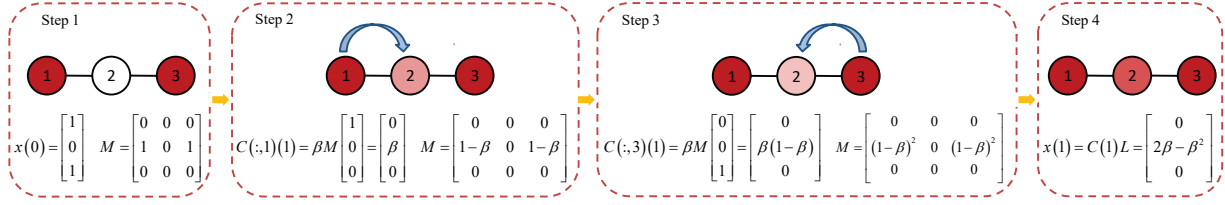}}
\caption{(Color online) An example network with 3 nodes, where node $1$ and node $3$ is the initial infected nodes. The probability of node $2$ $x_{2}(1)$ to be infected at time step 1 is $2\beta-\beta^{2}$ generated by the DMP method.}
\label{figure1}
\end{center}
\end{figure}

The spreading influence of the target node within a certain time $T^*$ is $\sum_{t=1}^{T^*}\sum_{j=1}^{n}x_{j}(t)$. During the spreading process, the DMP method solves the problem of nonlinear coupling by calculating the probability of the node to be infected by its infected neighbours sequentially via adjusting the state transition matrix $\mathbf{M}$. As shown in Fig.~\ref{figure1}, node $1$ and node $3$ are the initial spreaders. According to the DMP method, we firstly calculate the probability of node $2$ to be infected by node $1$, which is $\beta$. Then we update the state transition matrix $\mathbf{M}$. After that, we calculate the probability of node $2$ to be infected by node $3$, which is $\beta(1-\beta)$. And in the end, we get the probability of node $2$ to be infected by its both infected neighbours, which equals to $\beta+\beta(1-\beta)=1-(1-\beta)^2$. We note that this is the exact probability of the node $2$ to be infected. 

The DMP method could be extended to the SIS model with $\gamma=1$, where $\gamma$ is the probability of the infected nodes enter the susceptible state (see the details in the Data Analysis section). In SIS model, the difference is that at each time step $t$, the transition matrix $\mathbf{M}$ could be updated by  $m_{ij} = a_{ji}(1-x_{i}(t-1))$. For the temporal network, the network could be described by $\mathbf{A}(t)$ at each time step $t$. Thus, at time $t$ the transition matrix $\mathbf{M}$ could be updated by $m_{ij} = a_{ji}(t)(1-\sum_{r=0}^{t-1}x_{i}(r))$.


\section{Data Analysis}
\subsection{Data description}
We have tested the performance of DMP method in estimating the nodes' spreading influence according to the SIR and SIS models on four real networks. The first network is ``C. elegans", a directed network representing the neural network of Caenorhabditis elegans~\cite{watts1998collective}. The data is available at \url{https://snap.stanford.edu/data/C-elegans-frontal.html}. The second network is a scientific collaboration network, ``Erd\"{o}s", where nodes are scientists and edges represent the co-authorships.  The data can be freely downloaded from the web site \url{http://wwwp.oakland.edu/enp/thedata/}. The third one is an email communication network of University Rovira i Virgili (URV) of Spain, involving faculty members, researchers, technicians, managers, administrators, and graduate students~\cite{guimera2003self}. The data can be found at \url{http://konect.cc/networks/arenas-email/}. The last network is a directed network based on the ODLIS dictionary network. This a hypertext reference resource for library and information science professionals, university students and faculty, and users of all types of libraries. The node represents web site of Odlis and the edge represents the network are the connection between two web sites. This data is available at \url{http://networkdata.ics.uci.edu/netdata/html/ODLIS.html}. Basic statistical properties of these four networks are presented in Table~\ref{table1}.

Besides, we have also analyzed four real-world dynamic networks in order to evaluate the effectiveness of the DMP method. The first temporal network is \emph{Contacts in a workplace} (CW) network. This data includes contacts between individuals measured in an office building in France, from June 24 to July 3, 2013 \cite{Genois2015Data}. The second one is the \emph{Primary school} (PS) temporal network, where nodes are the children and teachers, and edges represent the contacts between them \cite{Gemmetto2014Mitigation,Stehle2011H}. The CW and PS network could be downloaded at \url{http://www.sociopatterns.org/datasets/}. The \emph{email-Eu-core-temporal-Dept1} (EM01) and \emph{email-Eu-core-temporal-Dept2} (EM02)~\cite{Paranjape2017M} temporal network are generated by using email data from a large European research institution, where edges present email between members of the research institution. These two datasets are available at \url{http://snap.stanford.edu/data/index.html}. In Table~\ref{table2}, we provide the detailed statistical properties of the above temporal networks.


\begin{table}[ht]
\begin{center}
\begin{tabular} {l c c c c}
  \hline \hline
    Network        &$n$      &$e$       & $\langle k\rangle$       & $1/\lambda_{1}$     \\ \hline
   C. elegans       &297      &2345      &7.896                            &0.109      \\
   Erd\"{o}s       &454      &1313      &5.784                            &0.079        \\
   Email           &1133     &5451      &9.622                            &0.048         \\
   Odlis           &2900     &18241     &6.290                            &0.077   \\
\hline \hline
\end{tabular}
\end{center}
\caption{Basic statistical features of C. elegans, Erd\"{o}s, Email, and Odlis networks, including the number of nodes $n$, the number of the edges $e$, the average degree $\langle k\rangle$ or $\langle k_{out}\rangle$ (for directed networks) and the reciprocal of the largest eigenvalue $1/\lambda_{1}$.}
\label{table1}
\end{table}

\begin{table}[ht]
\begin{center}
\begin{tabular} {l c c  }
  \hline \hline
    Network        &$n$      &$e$                  \\ \hline
   CW       &92      &1492                            \\
   PS       &242      &21295                                      \\
   EM01           &309     &11106                             \\
   EM02           &162     &7758                             \\
\hline \hline
\end{tabular}
\end{center}
\caption{Basic statistical features of CW, PS, EM01, and EM02 temporal networks, including the number of nodes $n$ and the number of the edges $e$ respectively.}
\label{table2}
\end{table}

\subsection{The SIR and SIS Model}

We apply the susceptible-infected-recovered (SIR) model~\cite{hethcote2000mathematics} and the susceptible-infected-susceptible (SIS)~\cite{pastor2001epidemic} model to simulate the spreading process and record the nodes' spreading influence at each time step. In the SIR model, there are three kinds of individuals: (i) susceptible individuals that could be infected, (ii) infected individuals which are able to infect their susceptible neighbors, and (iii) recovered individuals that will never be infected again. At each time step, every infected node will contact its neighbors and each of its susceptible neighbors will be infected with a probability $\beta$. Then the infected nodes enter the recovered state with a probability $\mu$. While in SIS model, there are only two kinds of individuals, i.e. the susceptible individuals and the infected ones. The infected nodes would infect its susceptible neighbors with the probability $\beta$ and enter the susceptible state with a probability $\gamma$. For single-node spreading, only one seed node is infected at the beginning, and all the other nodes are susceptible. While for multiple-nodes spreading, a set of nodes are infected and the rests are initially susceptible. At each time step $t$, the number of nodes that switch from the susceptible state to the infected state represents the node's spreading influence. For simplicity, we set $\mu=1$ in SIR model and $\gamma=1$ in SIS model. In this paper, all the analysis are based on the discrete-time dynamics.

\subsection{Kendall's Tau}
In this paper, we use the Kendall's tau to measure the correlation between the nodes' spreading influence and centrality measures (e.g., degree, eigenvector centrality and DMP method). For each node $i$, we denote $y_i$ as its spreading influence and $z_i$ as the target centrality measure, the accuracy of the target centrality in evaluating nodes' spreading influences can be quantified by the Kendall's Tau \cite{kendall1938new}, as
\begin{equation}
\tau=\frac{2}{\sqrt{(n(n-1)/2-n_{1})(n(n-1)/2-n_{2})}}\sum_{i<j}{\rm{sgn}}[(y_{i}-y_{j})(z_{i}-z_{j})],
\label{equation1}
\end{equation}
where $n_{1}=\sum_{i}{v_{i}(v_{i}-1)/2}$, $v_{i}$ is the number of the $i^{th}$ group of ties for the first quantity and $n_{2}=\sum_{j}{u_{j}(u_{j}-1)/2}$, $u_{j}$ is the number of the $j^{th}$ group of ties for the second quantity and sgn$(y)$ is a piecewise function: when $y>0$, sgn$(y)=+1$; $y<0$, sgn$(y)=-1$; when $y=0$, sgn$(y)=0$. $\tau$ measures the correlation between two ranking lists, whose value is between $[-1, 1]$ and a larger $\tau$ corresponds to a better performance.

\subsection{Numerical Result}

\begin{figure}
\begin{center}
\scalebox{0.9}[0.9]{\includegraphics{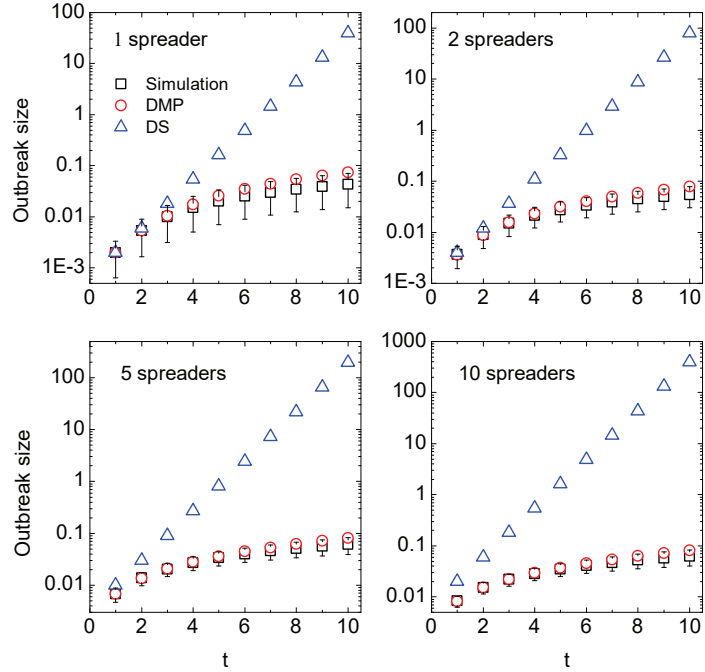}}
\caption{The performance of the DMP method and the DS centrality for evaluating the outbreak generated by both single spreader and multiple spreaders on regular network of $N=1000$ and $\langle k\rangle=20$ during the SIR spreading process with spreading rate $\beta=0.1$. The symmetric bars indicate the fluctuations around the average value computed on $10^5$ realizations of the stochastic process.}
\label{figure2}
\end{center}
\end{figure}

\begin{figure}[!ht]
\begin{center}
\scalebox{0.65}[0.65]{\includegraphics{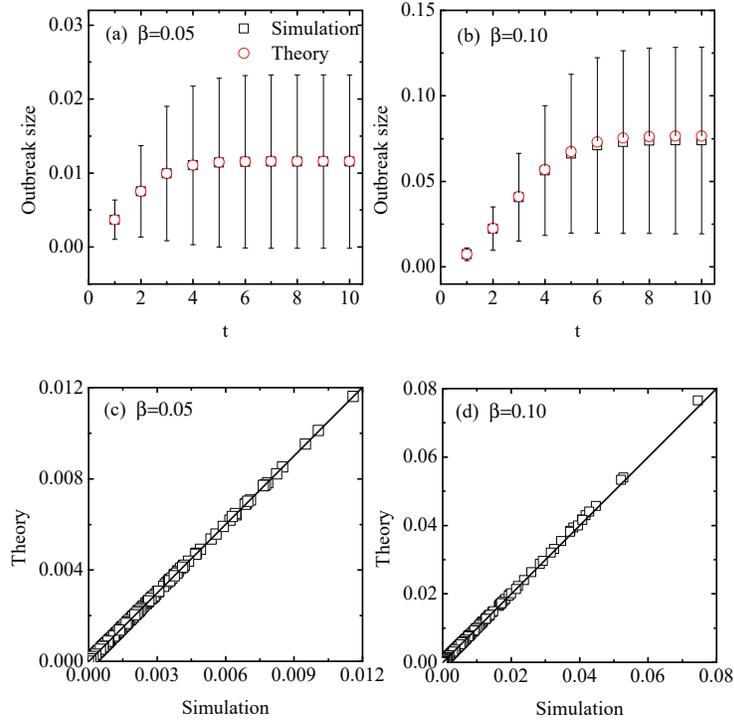}}
\caption{(Color online) The accuracy of the DMP method for evaluating nodes' spreading influence in a model network without loop of $N=500$ and $\langle k\rangle=8$ during the SIR spreading process. The subplot (a) and (b) show the  outbreak size of node 1 in model network generated by DMP method at each time steps when the spreading rate $\beta$ is 0.05 and 0.1 respectively. The subplot (c) and (d) show the outbreak size of all nodes in model network generated by the DMP method when spreading rate $\beta$ is 0.05 and 0.1 respectively. The symmetric bars indicate the fluctuations around the average value computed on $10^5$ realizations of the stochastic process.}
\label{figure2-1}
\end{center}
\end{figure}

\begin{figure}[!ht]
\begin{center}
\scalebox{0.75}[0.75]{\includegraphics{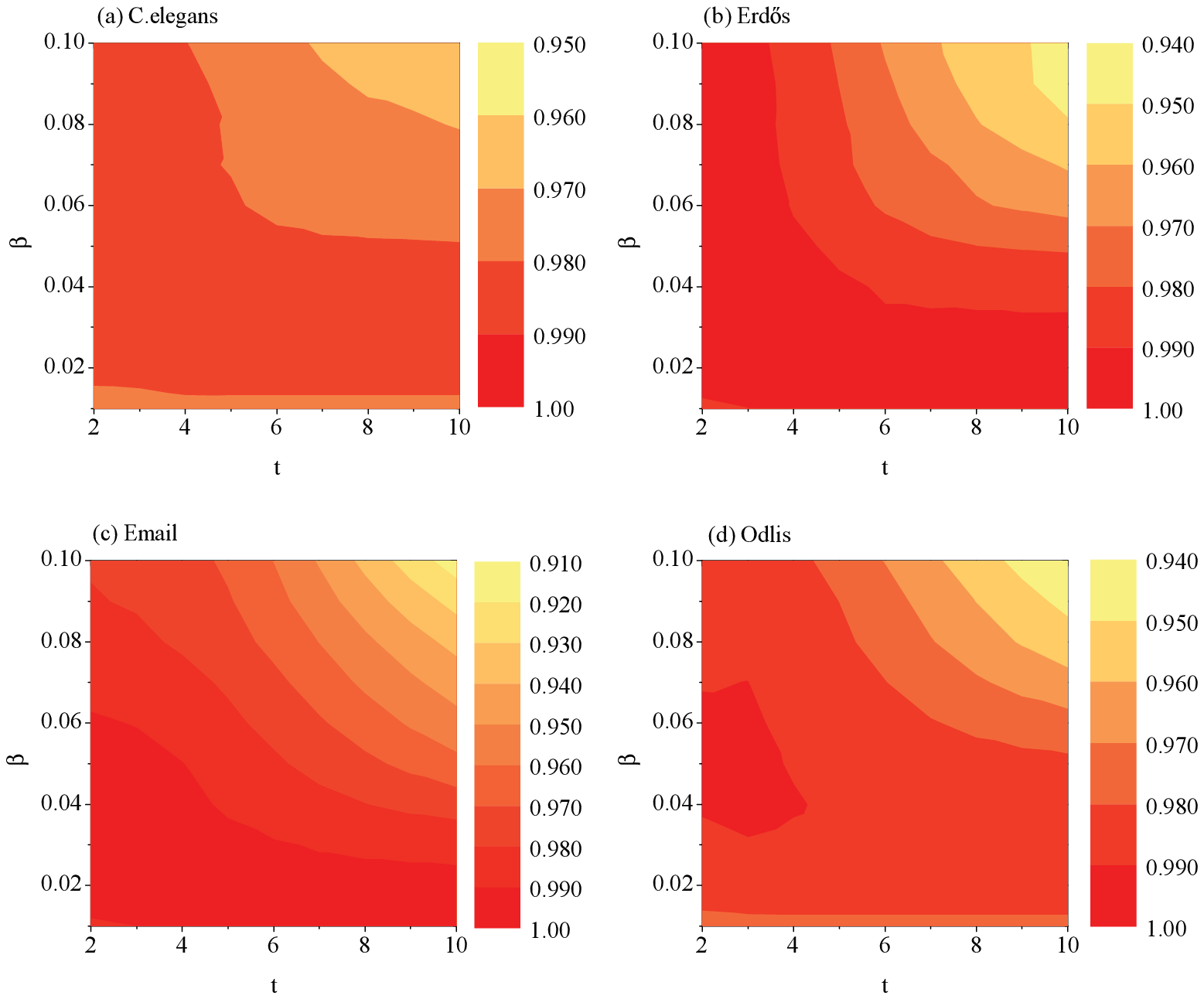}}
\caption{(Color online) The accuracy of the DMP method in evaluating nodes' spreading influences according to the standard SIR model in the four real networks, quantified by the Kendall's Tau. The spreading rate $\beta$ varies from 0.01 to 0.10, Each data point is obtained by averaging over $10^5$ independent runs.}
\label{figure3}
\end{center}
\end{figure}

\begin{figure}[!ht]
\begin{center}
\scalebox{0.55}[0.55]{\includegraphics{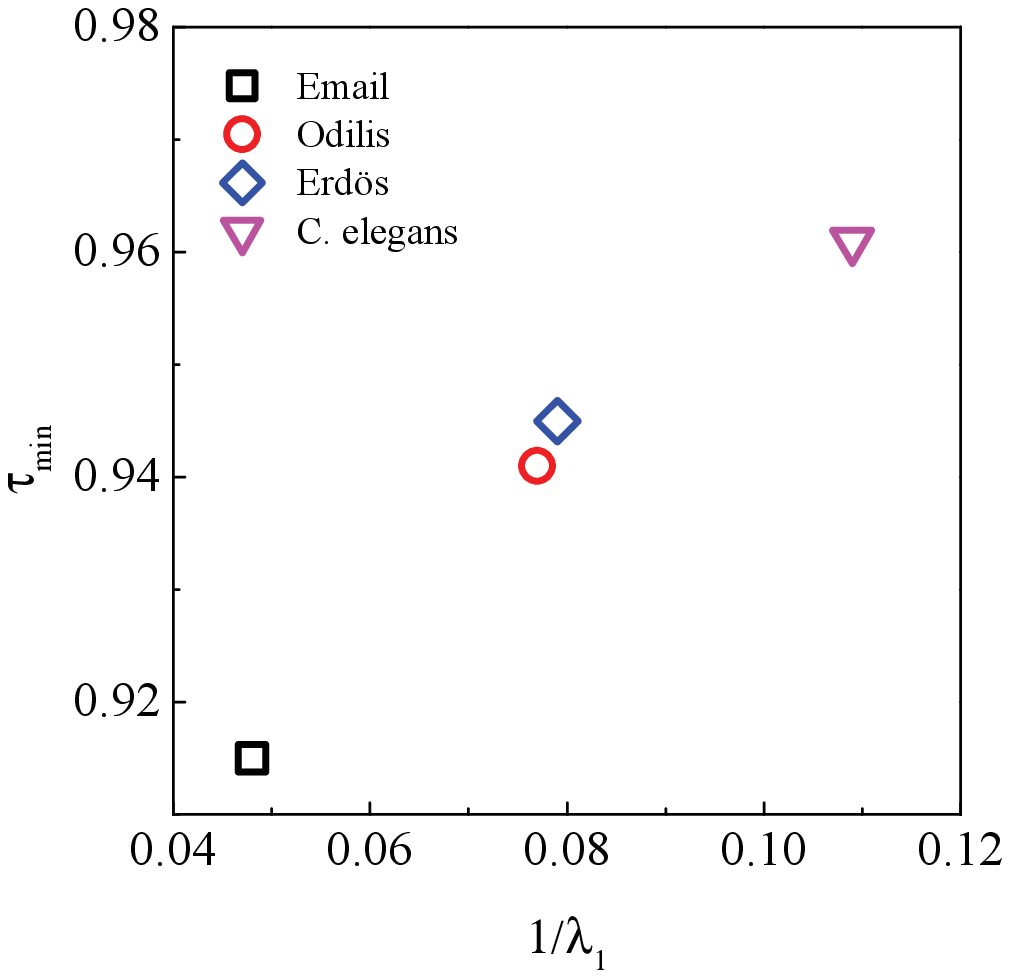}}
\caption{(Color online) The correlation between the minimum Kendall's tau $\tau_{min}$ and the inverse of the largest eigenvalue of the network $1/\lambda_{1}$.}
\label{figure4}
\end{center}
\end{figure}

Figure \ref{figure2} shows the comparison between the DMP method and the DS method for evaluating the nodes' spreading influence on four empirical networks. The results suggest that the DMP method could evaluate the outbreak size for both single and multi spreaders more accurately than the DS method. One could easily observe that the theoretical results generated by DMP method are over estimated. The main reason of this overestimation is that the nodes infected by the initial node would infect themselves when time steps $t\geq3$. One could then expect that a network without any loop would diminish this bias. As shown in Fig.\ref{figure2-1}, in a network without loop, the DMP method could evaluate the nodes' the spreading scope accurately compared with the simulation result on network.

We also test the performance of the DMP method in ranking nodes' spreading influence during the spreading process on SIR model with different spreading rates $\beta$. The spreading influence of an arbitrary node $i$ is quantified by the number of infected nodes and recovered nodes at $t$, where the spreading process starts with only node $i$ being initially infected. Here the Kendall's tau $\tau$ is used to evaluate the correlation between the nodes' spreading influence and the centrality measures (DMP method, degree and eigenvector centrality), where $\tau$ is in the range $\left[ -1,1 \right]$, and a larger value of $\tau$ indicates a better performance. As shown in Fig.~\ref{figure3}, in all the cases, the values of $\tau$ of the DMP method is always between 0.915 and 1.0, which suggests that the ranking lists generated by the DMP method is almost the same as the ones generated by the simulation result.

Furthermore, one can find that the accuracy of the DMP method for ranking nodes' spreading influence is affected by the network structure. As shown in Fig.~\ref{figure3}, the descent speed of $\tau$ in the C. elegans network are smaller than the ones in the Email network. To get deeper insight of how the network structure affect the performance of the DMP method, we analyze the correlation between the minimum Kendall's tau of the networks in Fig.~\ref{figure3} and the inverse of the largest eigenvalue of the network $1/\lambda_1$. The results are shown in Fig.~\ref{figure4}. With increasing value of  $1/\lambda_{1}$, the minimum Kendall's tau increases. For instance, in the C. elegans network, the reciprocal of the largest eigenvalue $1/\lambda_{1}$ is 0.109, which is significantly larger than that of the Email network (0.048). The fact that the DMP method performs particularly well in C. elegans network for ranking nodes' spreading influence indicates that the largest eigenvalue of the network is the main factor affecting the accuracy of the DMP method. A larger value of $1/\lambda_{1}$ would lead to a better performance of the DMP method.

\begin{figure}[!ht]
\begin{center}
\scalebox{0.75}[0.75]{\includegraphics{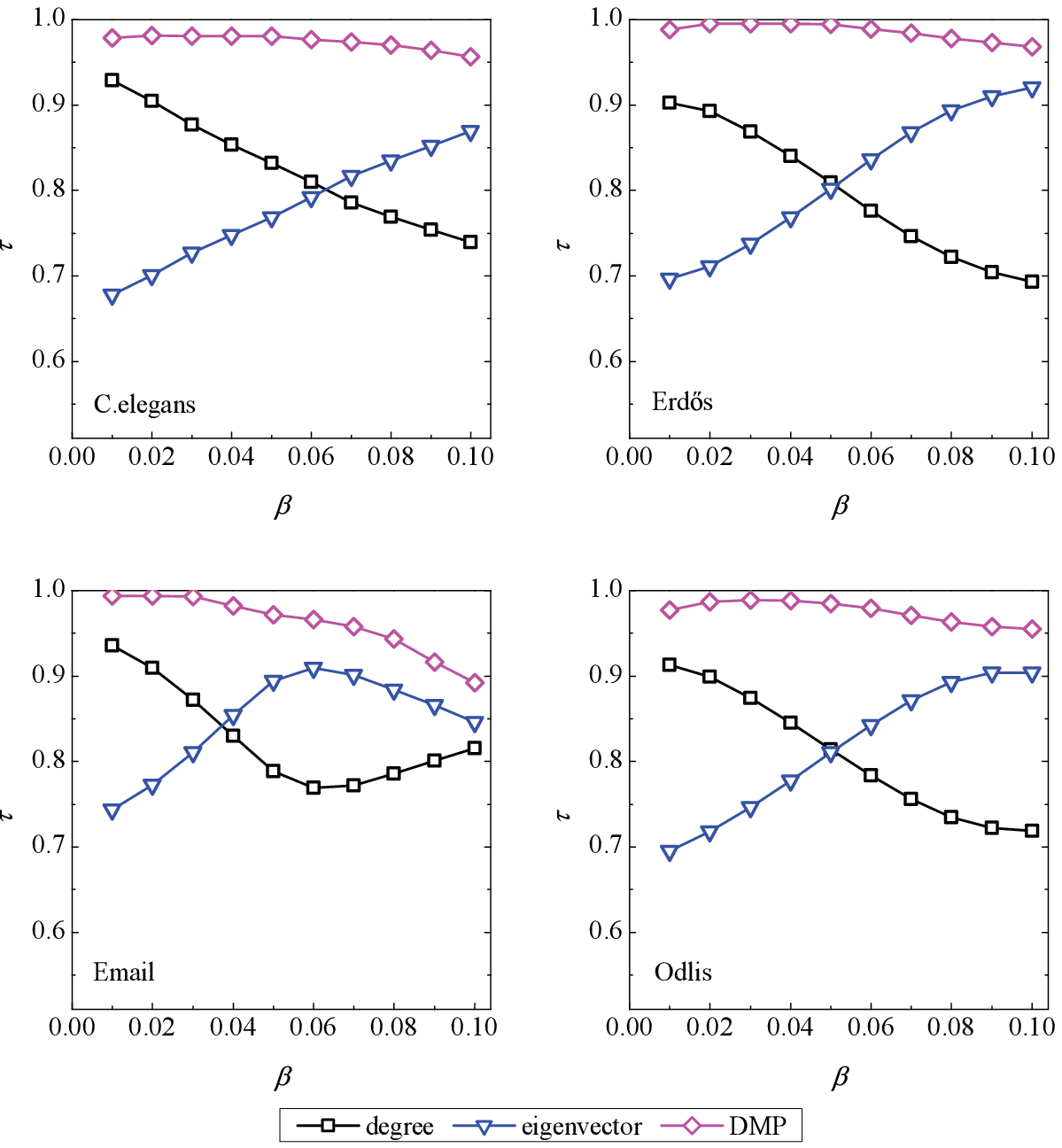}}
\caption{(Color online) The comparison among the DMP method, degree and eigenvector in evaluating nodes' spreading influences according to the standard SIR model with enough time steps until there is not any infected in the networks, quantified by the Kendall's Tau. The spreading rate $\beta$ varies from 0.01 to 0.10, Each data point is obtained by averaging over $10^5$ independent runs.}
\label{figure5}
\end{center}
\end{figure}

In Fig.\ref{figure3}, we have shown the spreading influence of nodes during the whole process. In the current experiment, we rank nodes' spreading influence in a special situation: the running time of the simulation is long enough such that there is not any infected node in the network. For each node, since we do not the exact convergent time step in the simulation, here we set a fix time step $T^{*}$ as 5 in the DMP method. The results are shown in Fig.~\ref{figure5}. The Kendall's tau $\tau$ of the DMP method is between 0.893 to 0.995, which indicates that the ranking lists generated by the DMP method and the SIR spreading process are almost the same. Compared with the degree and eigenvector centrality, the DMP method could locate the influential spreaders more accurately.

\begin{figure}
\begin{center}
\scalebox{0.73}[0.73]{\includegraphics{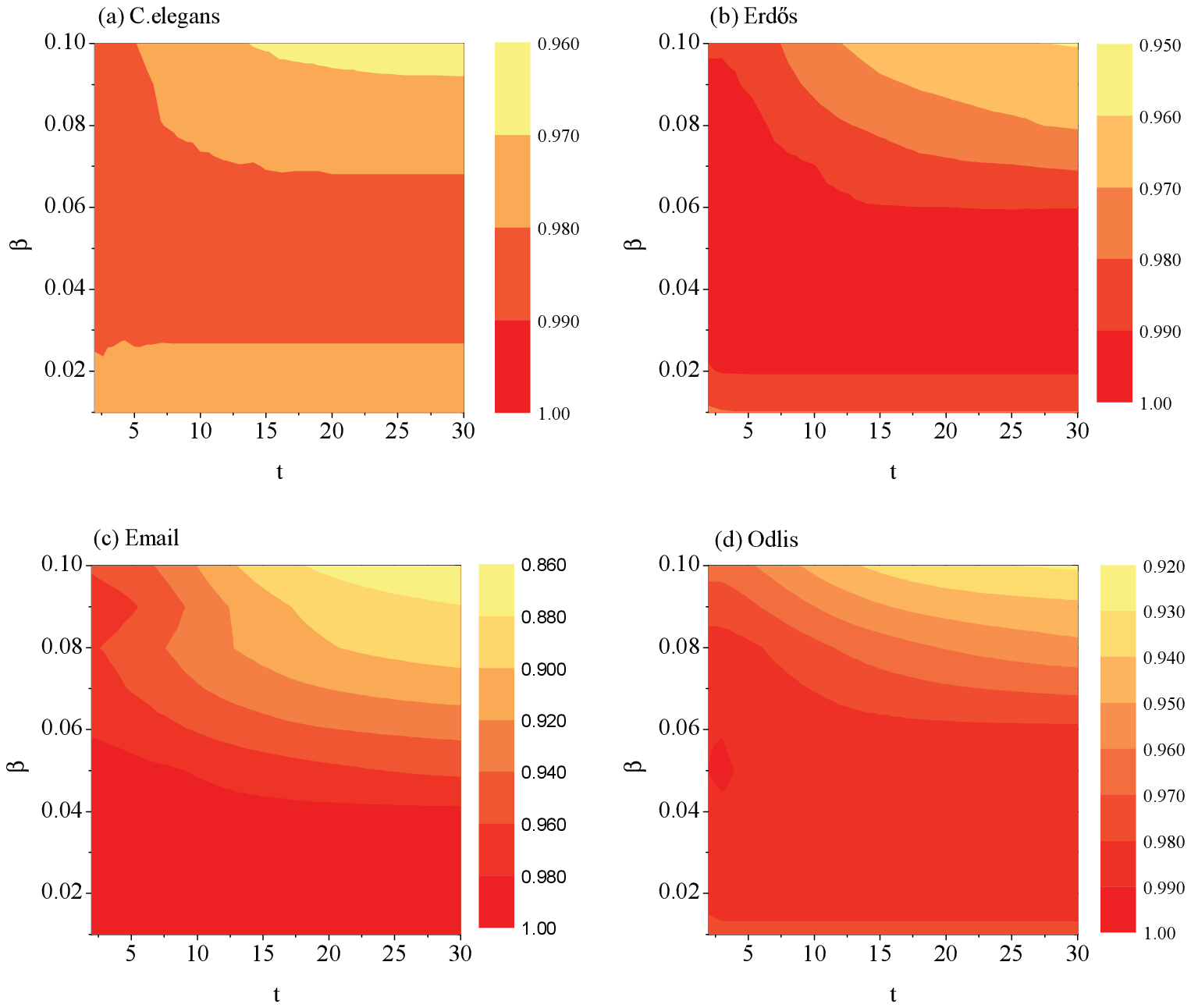}}
\caption{(Color online) The accuracy of the DMP method in evaluating nodes' spreading influences according to the standard SIS model in    four real networks, quantified by the Kendall's Tau. The spreading rate $\beta$ varies from 0.01 to 0.10, Each data point is obtained by averaging over $10^5$ independent runs.}
\label{figure6}
\end{center}
\end{figure}

\begin{figure}
\begin{center}
\scalebox{0.73}[0.73]{\includegraphics{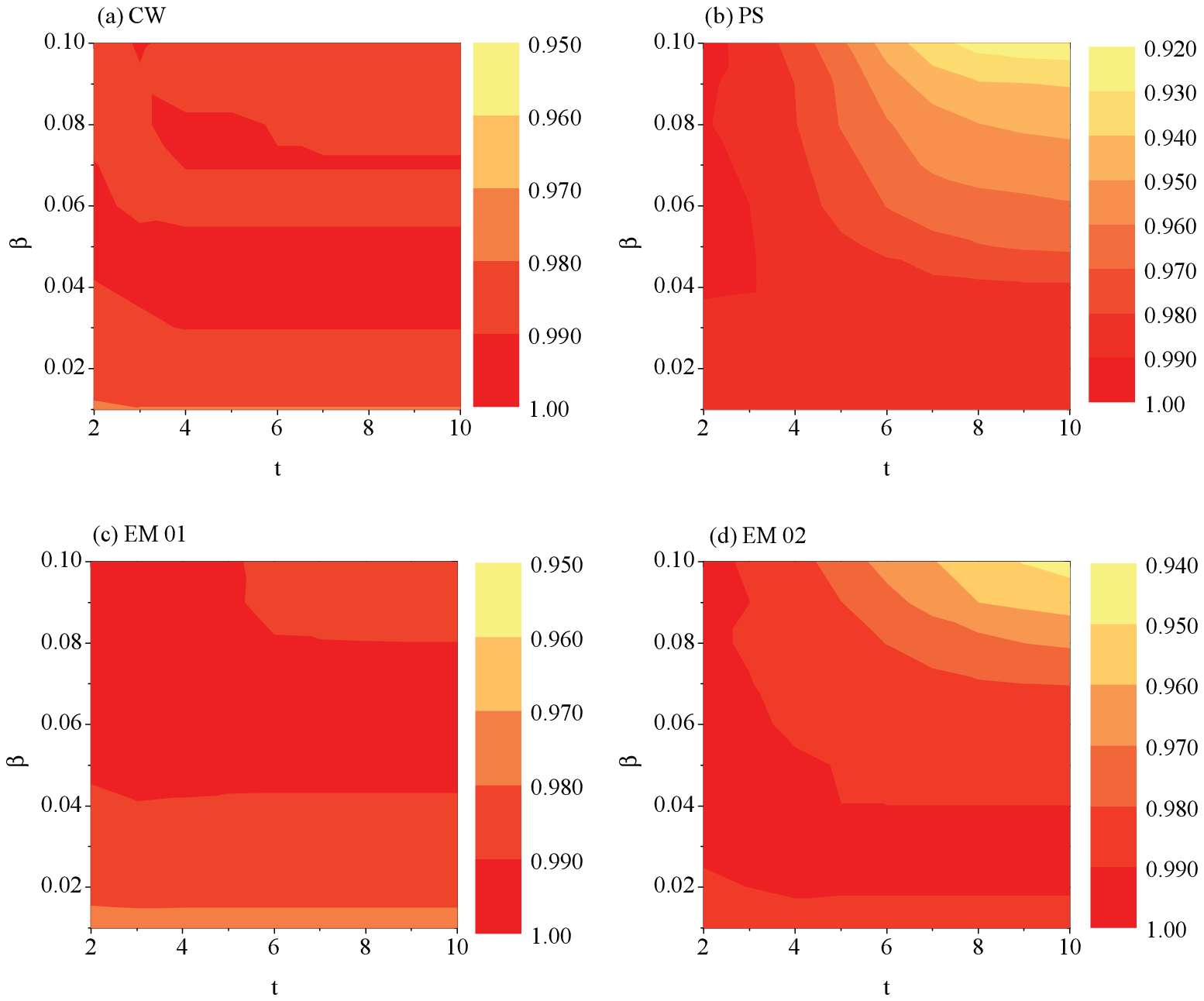}}
\caption{(Color online) The accuracy of the DMP method in evaluating nodes' spreading influences according to the standard SIR model in four real temporal networks, quantified by the Kendall's Tau. The spreading rate $\beta$ varies from 0.01 to 0.10, Each data point is obtained by averaging over $10^5$ independent runs.}
\label{figure7}
\end{center}
\end{figure}


The DMP method can also be used to evaluate the outbreak size in SIS model. For simplicity, we set $\gamma=1$ in SIS model. In this case, nodes in the network could be infected several times for high spreading rates $\beta$ and long time steps $t$. We set the final time step $t^{*}$ to 30. The results are shown in Fig. \ref{figure6}. The results are very similar to the ones in SIR model. The Kendall's tau $\tau$ of the DMP method is between 0.865 and 1.0. This indicates that the DMP method could also rank the nodes' spreading influence accurately in SIS model.

We have extended the DMP method to temporal networks. The results are shown in Fig.~\ref{figure7}. The Kendall's tau $\tau$ is between 0.923 to 0.992, which indicates that the ranking lists generated by the DMP method and the real SIR model on temporal are highly identical to each other. Therefore, the DMP method could be used to detect the influential nodes in temporal network accurately.

\section{Discussions}
The essential question of ranking the node spreading influence is how to estimate the outbreak size of the initial spreader\cite{chen2019identifying,bucur2019beyond}. To answer this question, one needs to fix the nonlinear coupling issue during the spreading process. In this paper we present a new method to evaluate the spreading scope from the perspective of Markov chain process, namely dynamics Markov process (DMP). This method solves the problem of nonlinear coupling by adjusting the state transition matrix, in which the elements of the matrix are the probabilities of nodes in susceptible state. The simulation results show that the DMP method could estimate the nodes' spreading scope at each time steps accurately in directed network without loop. Furthermore, according to the empirical results on four real networks, for both the SIR and SIS model, the ranking list generated by the DMP method is very close to the the ones of the simulation results, especially when the spreading rate and time step is small.

The DMP method could also be used to evaluate the nodes' spreading scope generated by multi-spreaders. Our simulation results indicate that when there exist multiple spreaders, the DMP method significantly outperforms the DS\cite{liu2016locating} centrality with increasing values of spreaders and time steps. The key to identifying  multiple influential spreaders is to solve the overlap problem~\cite{wang2017identifying}, which is the non-linear couple problem during the spreading process. Given the fact that the DMP method is a non-linear method, it will be able to identify multiple influential spreaders by using the greedy approach~\cite{altarelli2014containing,morone2015influence,guo2016identifying}. Moreover, the DMP method is also suitable for detecting the influential nodes in dynamic networks.

Comparing to the other methods in evaluating the outbreak size of the spreading dynamics, e.g., the Message-Passing Techniques \cite{altarelli2014containing} and the Percolation~\cite{hu2015quantify}, the DMP method evaluates the spreading scope from the perspective of Markov process, and provides a general framework for ranking node spreading influence. Therefore. it can be extended and applied in modeling many other important dynamics such as Ising model~\cite{dorogovtsev2008critical}, Boolean dynamics~\cite{kaufmann1993origins}, voter model~\cite{castellano2009statistical}, synchronization~\cite{arenas2008synchronization}, and so on.

\section{Authors contributions}

Jian-Hong Lin and Zhao Yang performed the analysis. Jian-Hong Lin and Claudio J. Tessone designed the research. All authors wrote, reviewed and approved the manuscript.

\section{Acknowledgements}
This work is partially supported by the China Scholarship Council and the Swiss National Science Foundation grant \#200021\_182659.
\bibliography{Bibliography}

\end{document}